\documentclass[journal,article,submit,moreauthors,pdftex]{article}
\usepackage[utf8]{inputenc}

\usepackage{listings}
\usepackage{color}
\usepackage{graphicx}
\usepackage{hyperref}
\usepackage{authblk}

\definecolor{dkgreen}{rgb}{0,0.6,0}
\definecolor{gray}{rgb}{0.5,0.5,0.5}
\definecolor{mauve}{rgb}{0.58,0,0.82}

\date{\today}
\begin{document}

\title{Forensic Log-Based Detection for Keystroke Injection BadUSB Attacks}
\maketitle
\author{Georgios Karantzas, BSc Student, Department of Informatics, University of Piraeus, Greece}

\begin{abstract}
	This document describes an experiment with main purpose to detect BadUSB attacks that utilize external
 HID (Human Interaction Device) "gadgets" to inject keystrokes and acquire remote code execution. One of the main goals, is to detect such activity based on behavioral factors and allow everyone with a basic set of cognitive capabilities ,regardless of the user being a human or a computer, to identify anomalous speed related indicators but also correlate such speed changes with other elements such as commonly malicious processes like "powershell" processes being called in close proximity timing-wise, PnP device events occurring,and long keyboard "sleeps" along with "Lock Key" abuse correlated with driver images loaded. 
\end{abstract}

\section{Introduction}
For our detection purposes, we need to consume events that will be coming directly from low level components of the Windows OS. We decided to take into consideration older publications such as abuse of USB2 and USB3 ETW providers and leverage such a provider as a "keylogger" (\textit{Microsoft-Windows-USB-USBPORT and Microsoft-Windows-USB-UCX}). Due to technical reproduction issues, however, and due to the facts that logging to a file would be synchronous and direct but also after seeing that some amount of customization would be needed capture-wise, we decided to utilize upper filters for one of the main parts of the proof-of-concept implementation. Those keyboard captures should be accompanied by a timestamp to give us an idea of how far they are from each other and whether the keystroke ratio represents a human user. The second part, can easily be accomplished via ETW barring its asynchronicity, buffer-based system limitations such as flushing logs all together along with mixing of events, that will not cause any issue for us as long as timestamps and data are not corrupt. A final note to keep in mind with such encounters is that privacy of the data need to be taken seriously, in our case we only need to know if the key was pressed and if it was a "Lock" key, if it wasn't it shouldn't be transmited through file writes or ETW from the driver to usermode.A past example to avoid was HP Synaptics keylogger distribution. Finally, cognitive capabilities will be needed given the product of the POC will be log-related and through looking at it as a timeline, one will understand the anomalies between a human typing normally speed-wise and a device typing abnormally fast and launching potentially malicious processes and components.

\section{Malicious BadUSB Attacks Today}
The most well-known BadUSB attack vector is probably the commercial "RuberDucky" which initially started as a sysadmin gadget to automate mundane tasks. This platform evolved into the most notorious attacker gadget with a series of community-backed payloads whose main capabilities include dual usage as a USB Stick, data exfiltration, HID Interaction (Keystroke Injection) and even its own scripting language, DuckyScript. This is backed by a USB2.0 hardware interface and support for USB-c. Some of the most advanced features include, copying payload to itself, "OUT endpoint" usage via "Lock Keys" "spamming", Keystroke Reflection and even features like VendorID and ProductID spoofing. 
For the purposes of this experiment, our "go-to" tool is going to be the RubberDucky's latest version as of this writing and various payloads will be employed across all our experiments but mainly, one launching powershell to dump credential files. In  most of the real-life cases powershell or other LoLBins will be used to run code and files will be dropped and executed. We should keep in mind the evasion features as they may avoid "hardcoded" detections but also in the case of sleeps, abnormally alter the keystroke timeline and introduce a cognitive anomaly.

\section{Event Tracing for Windows and its Drawbacks}
As discussed previously, one of our main source of information will be ETW. Initially introduced as a debugging feature, ETW gained a lot of attention due to the facts that it was easily usable, had a large amount of default and 3-rd party providers, Microsoft Introduced Patch-Guard-compliant Kernel API hooks with it and in general could provide easily vast amounts of telemetry from usermode and kernelmode providers.
Although it may sound tempting, this mechanism is simply not a silver bullet for all kinds of detections and telemetry ingestion.
Below you can find a table of the drawbacks and positive aspects of ETW.

\begin{center}
  \begin{tabular}{ l | c | r }
    \hline
    \textbf{Pros} & \textbf{Cons} \\ \hline
    Vast Amount Of Telemetry Flavors & Buffer Based       \\ \hline
    API Monitoring & Buffers May Flood and Degrade Consumption       \\ \hline
    Pre-structured Network Monitoring & Asynchronous      \\ \hline
    Easy to Deploy and Consume & Events May and Will be Missed by Design       \\ \hline
    Callstacks Provided & Timing Attacks \\ \hline
    - & Out of Order Events \\ \hline
    - & Performance Overhead \\ \hline
    \hline
  \end{tabular}
\end{center}

To further elaborate, we should consider a hypothetical example of what is "bad" usage of the mechanism.
Supposing we would like to monitor local memory modifications such as allocation and re-protections and even scanning memory for possibly malicious patterns and PIC (position independent code), below are a few empirical consequences faced when dealing with real-life production that can reduce effectiveness and make one's life more difficult.
We should keep in mind that for such an experiment ETWTI is used.

\begin{itemize}
  \item Reading from a process that exited and its PID (Process Identified) was re-issued as the event came "late".
  \item Attempting to find a module in-memory that is not already loaded or was removed during a short timeframe, excluding NTDLL which is a standardized case due to both its nature of existing always and Known DLLs behavior in terms of position. This whole situation may cause access violations given the unstable and non-synchronous way loaded modules are kept track of.
  \item Performance impact and overwhelming produced load of data to be analyzed given APIs monitored may be used by the Windows loader and cause massive overhead and difficulties.
  \item Applications that reside inside the main application such as additional plugins and anti-exploit agents may increase overhead even more.
\end{itemize}

Summing up the purposes of the mechanism, one should think twice before selecting ETW for its task by considering whether timely fashion of the event processing is of critical importance to them.
Regarding the practical implementation, in order to avoid needless development overhead and also avoid APIs like "Tdh*" etc. , we will reside to using "KrabsEtw", which provides a slick OOP wrapper around ETW libraries and allows for easy consumption of events form various providers, including the trace sessions of the "Kernel Logger" and also parsing them effectively along with special properties that may need to be passed such as obtaining stacks.

\section{Upper Filters for Keyboards}
Keyboard keyloggers are a very classical example of introductory projects in kernel development. The are various paths one could follow to achieve the results wanted. Quite interestingly we started from a specific approach and transitioned to another more intricate one to achieve higher kevels of functionality, stability and effectiveness. The initial approach was setting an upper filter device and a completion routine that will intercept and log keyboard "MakeCodes" from "PKEYBOARD\_INPUT\_DATA" structure pointers. The code will hook also on some other IRP Dispatch routines to ensure overall proper functionality. We should note that the code is non-PnP so far and target to filter kbdclass, the class driver under which all kinds of port-mini-port driver pairs exist for keyboards, regardless of their type (PS\\2 , USB etc). Also, a bit of waiting will happen during the unload routine to ensure all IRPs were processed. However, for both of the driver cases, unloads are very basic and don't support PnP which may and will result to inconsistencies. Setting of the interception completion routine happens by the "IoSetCompletionRoutine" call inside the Dispatch Routine for IRP\_MJ\_READ. In general, such a non-PnP aware filter driver may result in various issues, therefore, an alternative was chosen. This alternative included utilizing a KMDF driver inside which "listens" to the PnP manager for new devices. It is worth noting that one should register this filter as an "UpperFilter" above "kbdclass" inside the registry under the appropriate GUID. What you will see in the code below is the code setting the main hooking routine, after the appropriate callbacks and data were setup and upon calling WDF\_DRIVER\_CONFIG\_INIT to listen for devices, WDF\_OBJECT\_ATTRIBUTES initializer and even after having set the WDF Driver device.
In the code below, we create a device with our extension data.

\begin{lstlisting}
        WdfFdoInitSetFilter(DeviceInit);
        WdfDeviceInitSetDeviceType(DeviceInit, FILE_DEVICE_KEYBOARD);

        WDF_OBJECT_ATTRIBUTES_INIT_CONTEXT_TYPE(&wdfAttrib, USB_TROLL_EXTENSION);
        status = WdfDeviceCreate(&DeviceInit, &wdfAttrib, &hControlDevice);
        if (!NT_SUCCESS(status))
        {
            __leave;
        }
        
        WDF_IO_QUEUE_CONFIG_INIT_DEFAULT_QUEUE(&ioQueueConfig,
            WdfIoQueueDispatchParallel);

        ioQueueConfig.EvtIoInternalDeviceControl = CbEvtIoInternalDeviceControl;

        status = WdfIoQueueCreate(hControlDevice,
            &ioQueueConfig,
            WDF_NO_OBJECT_ATTRIBUTES,
            WDF_NO_HANDLE
        );
\end{lstlisting}

To simply put what will follow, when a new device is added, we will intercept it and attach ourselves as a filter, then using a modified WDF\_IO\_QUEUE\_CONFIG structure with our own set PFN\_WDF\_IO\_QUEUE\_IO\_INTERNAL\_DEVICE\_CONTROL and by creating an I\\O queue through WdfIoQueueCreate, we essentially get a foothold so we can use "WdfRequestRetrieveInputBuffer" and maybe appropriate driver contexts and finally hook onto CONNECT\_DATA's ClassService routine with "ServiceCallbackDummy".
A function driver calls the class service callback in its ISR dispatch completion routine. The class service callback transfers input data from the input data buffer of a device to the class data queue. These conditions make it a prime hooking target.
You can see below the code responsible for placing the hooking routines themselves.
More specifically, we change the device and service callback to our own.

\begin{lstlisting}[language=Python]
_Function_class_(EVT_WDF_IO_QUEUE_IO_INTERNAL_DEVICE_CONTROL)
_IRQL_requires_same_
_IRQL_requires_max_(DISPATCH_LEVEL)
VOID
CbEvtIoInternalDeviceControl(
    _In_ WDFQUEUE      Queue,
    _In_ WDFREQUEST    Request,
    _In_ size_t        OutputBufferLength,
    _In_ size_t        InputBufferLength,
    _In_ ULONG         IoControlCode
)
{
    BOOLEAN bRetSuccess = TRUE;
    WDF_REQUEST_SEND_OPTIONS options;
    WDFDEVICE hDevice;
    NTSTATUS status = STATUS_UNSUCCESSFUL;
    CUSTOM_EXTENSION    pData = NULL;
    PCONNECT_DATA connectData = NULL;
    size_t length;

    UNREFERENCED_PARAMETER(OutputBufferLength);
    UNREFERENCED_PARAMETER(InputBufferLength);
    UNREFERENCED_PARAMETER(Request);

    hDevice = WdfIoQueueGetDevice(Queue);
    pData = GetData(hDevice);

    switch (IoControlCode) 
    {
    case IOCTL_INTERNAL_KEYBOARD_CONNECT:

        if (pData->UpperConnectData.ClassService != NULL) 
        {
            status = STATUS_SHARING_VIOLATION;
            break;
        }

        status = WdfRequestRetrieveInputBuffer(Request,
            sizeof(CONNECT_DATA),
            &connectData,
            &length);
        if (!NT_SUCCESS(status)) {
            break;
        }
        NT_ASSERT(length == InputBufferLength);

        pData->UpperConnectData = *connectData;

        connectData->ClassDeviceObject = WdfDeviceWdmGetDeviceObject(hDevice);
        connectData->ClassService = ServiceCallbackDummy;

        break;
    case IOCTL_INTERNAL_KEYBOARD_DISCONNECT:
        break;
    default:
        break;
    }
  

    WDF_REQUEST_SEND_OPTIONS_INIT(&options,  WDF_REQUEST_SEND_OPTION_SEND_AND_FORGET);
    bRetSuccess = WdfRequestSend(Request, WdfDeviceGetIoTarget(hDevice), &options);
    if (!bRetSuccess ) 
    {
        status = WdfRequestGetStatus(Request);
        WdfRequestComplete(Request, status);
    }

    return;
}
\end{lstlisting}

Below you can find the relevant callback we spoofed previously so we can log keystrokes via the "MakeCode" intercepted through the buffer that was being passed through the stack.
The concept behind the code is forwarding to the next driver after we have en-queued a "SafeLog" routine, that will take all precautions needed to log to our file the keyboard code safely in a multi-threaded environment.

\begin{lstlisting}

VOID
DummyServiceCallback(
    _In_ PDEVICE_OBJECT  DeviceObject,
    _In_ PKEYBOARD_INPUT_DATA InputDataStart,
    _In_ PKEYBOARD_INPUT_DATA InputDataEnd,
    _Inout_ PULONG InputDataConsumed
)
{
    CUSTOM_EXTENSION   pData;
    WDFDEVICE   hDevice;
    PKEYBOARD_INPUT_DATA pInputData;
    hDevice = WdfWdmDeviceGetWdfDeviceHandle(DeviceObject);

    pUData = CustomGetData(hDevice);

    for (pInputData = InputDataStart; pInputData != InputDataEnd; pInputData++)
    {
        TpkEnqueueWorkItem(&gDrv.ThreadPool,
            SafeLog, 
            SafeLog,
            (PVOID)pInputData->MakeCode);
    }

    (*(PSERVICE_CALLBACK_ROUTINE)(ULONG_PTR)pData->UpperConnectData.ClassService)(
        pData->UpperConnectData.ClassDeviceObject,
        InputDataStart,
        InputDataEnd,
        InputDataConsumed);
}

\end{lstlisting}
To sum up, this stealthier, lesser common and possibly more "hacky" hooking approach was employed to increase the chances of the keyboard monitoring driver of being more "universal" and "stable" accross all kind of Windows OS whether on a Virtual Machine or a physical machine, unlike the predecessor.

\section{Microsoft-Windows-Kernel-Process and Its Use}
We decided to utilize the aforementioned ETW provider to collect information about Image Loads (incl. Drivers) and Process Creation events. We limited ourselves to only these event categories.
Below you can find some example code of adjusting and enabling providers on a certain trace session which will act as a glue between the consumer and the provider.

\begin{lstlisting}
LoggerSession::LoggerSession(
	_In_ krabs::c_provider_callback CbHandler,
	_In_ bool EnableProcess) :
	BadUsbTrace(L"BadUsbTrace"),
	PrcProvider(L"Microsoft-Windows-Kernel-Process")
{
	EnableProc = EnableProcess;
	
	if (EnableProc)
	{
		PrcProvider.any(0x50); 
		PrcProvider.add_on_event_callback(CbHandler);
		BadUsbTrace.enable(PrcProvider);
	}
}
\end{lstlisting}

We are interested into writing easily parse-able logs that will include timestamps, names of images and process IDs when applicable. We utilize KrabsETW's parsing capabilities and created a callback handler to assist us with the process of creating logs.
The code below could serve as a simplistic example of how we chose to handle the event data.
The "FileLog" class will create and synchronously lock the file where we will output our logs as well as handle the text processing.
We use KrabsEtw's default parser code to "excavate" event data we need and assign them to variables with proper initial values.
We finally decide to work with a "std::wstring" and write our log from a C-string format.

\begin{lstlisting}
void ProcHandler::CbProcHandler(
	_In_ const EVENT_RECORD& Record,
	_In_ const krabs::trace_context& TraceContext)
{
	krabs::schema ProcSchema(Record, TraceContext.schema_locator);
	krabs::parser ProcParser(ProcSchema);

	const wchar_t* taskProcCreate  = L"ProcessStart";
	const wchar_t* taskProcLoadImg = L"ImageLoad";

	FileLogger FileLog(L"imagesprocs.txt");

	if (wcscmp(ProcSchema.task_name(), taskProcCreate) == 0) 
	{
		uint32_t procId = 0;
		ProcParser.try_parse(L"ProcessID", procId);

		if (!procId)
		{
			return;
		}

		FILETIME createTime = { 0 };
		ProcParser.try_parse(L"CreateTime", createTime);

		LARGE_INTEGER createTimeLInt = { 0 };
		createTimeLInt.HighPart = createTime.dwHighDateTime;
		createTimeLInt.LowPart  = createTime.dwLowDateTime;

		if (!createTimeLInt.QuadPart)
		{
			return;
		}

		std::wstring imageName = { 0 };
		ProcParser.try_parse(L"ImageName", imageName);

		if (!imageName.size())
		{
			return;
		}

		std::wstring timeStamp  = std::to_wstring(createTimeLInt.QuadPart);
		
		std::wstring procIdWstr = std::to_wstring(procId);

		std::wstring logStr = L"ProcessLog:" + timeStamp + L":" + procIdWstr + L":" + imageName + L":EndProcessLog\n";

		FileLog.LogToFile(logStr.c_str());
	}
\end{lstlisting}

Based on how we architected this initial provider, we decided to extend the architecture but maintain the same backbone with additional providers or features we may wanna add. 
An example graph of the architecture is provided below.
Essentially, our "ETW Logger Manager Class" sets the appropriate callbacks and event limitations that will handle data and lo them appropriately using the "File Logger Class".

\includegraphics{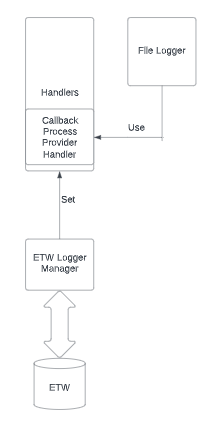}

Finally , we shall provide some raw output from the tool to give the reader an idea of what to expect.
The "ImageLog" entries represent the output received when a Razer mouse was attached and the "ProcessLog" entries represent random process executions as an example.

\begin{lstlisting}
ImageLog:2968525321:\Device\HarddiskVolume3\Windows\System32\drivers\hidusb.sys:EndImageLog
ImageLog:1597723157:\Device\HarddiskVolume3\Windows\System32\drivers\RzDev_0084.sys:EndImageLog
ImageLog:1659522856:\Device\HarddiskVolume3\Windows\System32\drivers\RzCommon.sys:EndImageLog


ProcessLog:133197924344035078:8884:\Device\HarddiskVolume3\Windows\
System32\dllhost.exe:EndProcessLog


ProcessLog:133197924344358867:17944:\Device\HarddiskVolume3\Windows\
System32\Taskmgr.exe:EndProcessLog


ProcessLog:133197924344422757:15076:\Device\HarddiskVolume3\Windows\
System32\consent.exe:EndProcessLog


ProcessLog:133197924344814263:11828:\Device\HarddiskVolume3\Windows\
System32\Taskmgr.exe:EndProcessLog


ProcessLog:133197924349657313:16168:\Device\HarddiskVolume3\Program Files\Rivet Networks\SmartByte\RAPS.exe:EndProcessLog


ProcessLog:133197924364657393:13608:\Device\HarddiskVolume3\Windows\
System32\cmd.exe:EndProcessLog

ProcessLog:133197924364676889:15436:\Device\HarddiskVolume3\Windows\
System32\conhost.exe:EndProcessLog
\end{lstlisting}

\section{Microsoft-Windows-Kernel-PnP and Its Use}
The aforementioned provider, contains various events related to the PnP system and will allow us to have some extra information, or visibility if you will. Output by itself may not be self-explanatory and even chaotic at a mass scale, but its final goal is to be used in conjunction with other metrics in a log correlation process where its value will increase.
Below you can see a part from a generic log produced by the attachment of a Razer mouse, this will provide the reader with an idea of what to expect as output. 

\begin{lstlisting}
PnpLog:USB\VID_1532&PID_0084\5&1c5b639f&0&2:133197915652312694:EndPnpLog
PnpLog:USB\VID_1532&PID_0084&MI_00\6&34f4fee0&0&0000:133197915652451881:EndPnpLog
PnpLog:USB\VID_1532&PID_0084&MI_01\6&34f4fee0&0&0001:133197915652492370:EndPnpLog
PnpLog:USB\VID_1532&PID_0084&MI_02\6&34f4fee0&0&0002:133197915652554801:EndPnpLog
PnpLog:USB\VID_1532&PID_0084&MI_03\6&34f4fee0&0&0003:133197915652586085:EndPnpLog
PnpLog:HID\VID_1532&PID_0084&MI_00\7&f76681d&0&0000:133197915652597149:EndPnpLog
PnpLog:HID\VID_1532&PID_0084&MI_00\7&f76681d&0&0000:133197915652597324:EndPnpLog
PnpLog:HID\VID_1532&PID_0084&MI_01&Col01\7&334da5df&0&0000:133197915652599115:EndPnpLog
PnpLog:HID\VID_1532&PID_0084&MI_01&Col01\7&334da5df&0&0000:133197915652599264:EndPnpLog
PnpLog:HID\VID_1532&PID_0084&MI_01&Col02\7&334da5df&0&0001:133197915652600646:EndPnpLog
PnpLog:HID\VID_1532&PID_0084&MI_01&Col03\7&334da5df&0&0002:133197915652605277:EndPnpLog
PnpLog:HID\VID_1532&PID_0084&MI_01&Col04\7&334da5df&0&0003:133197915652609096:EndPnpLog
PnpLog:HID\VID_1532&PID_0084&MI_01&Col05\7&334da5df&0&0004:133197915652613813:EndPnpLog
PnpLog:RZVIRTUAL\VID_1532&PID_0084&MI_00&Col03\7&334da5df&0&01:133197915652618792:EndPnpLog
PnpLog:HID\VID_1532&PID_0084&MI_02\7&1b8a199a&0&0000:133197915652629790:EndPnpLog
PnpLog:HID\VID_1532&PID_0084&MI_02\7&1b8a199a&0&0000:133197915652629984:EndPnpLog
PnpLog:HID\VID_1532&PID_0084&MI_03\7&3c68d55&0&0000:133197915652633449:EndPnpLog
PnpLog:RZCONTROL\VID_1532&PID_0084&MI_00\8&e070abb&0:133197915652692960:EndPnpLog
PnpLog:RZCONTROL\VID_1532&PID_0084&MI_00\8&e070abb&0:133197915652692976:EndPnpLog
PnpLog:HID\VID_1532&PID_0084&MI_00&Col03\8&16f5acbd&0&0000:133197915652697204:EndPnpLog
\end{lstlisting}

There is no further need to provide code as the architecture was already discussed.

\section{Forensically Detecting the "Rubber Ducky"}

For our scenario, we utilized a publicly available "Ruber Ducky" script that would dump credentials and exfiltrate them right after.

As in every forensic analysis, we would need some indicators allowing us to initial identify footprints that will lead to a correlation complex unraveling the attack's steps. To start such a procedure, we will need to first have a full view of the raw data we can dig through. So far we have the following sets of data:

\begin{itemize}
  \item Timestamp-infused logs of keyboard usage.
  \item Timestamp-infused logs of spawned processes.
  \item Timestamp-infused logs of drivers loaded.
  \item Timestamp-infused logs of PnP devices loaded.
\end{itemize}

In some cases, we could try directly finding continuous "lock key" presses across the log file. In our case, such an attack is out-of-scope, therefore we will attempt following a methodologically diferent procedure. 
Investigating the final two seemed like a more friendly approach towards human investigators. We will want to note the fact that close to the loading of \textbf{hidusb.sys}  the HID-related drivers we can identify both \textbf{powershell.exe} presence and a PnP device named \textbf{"ATMEL"}.

The later device, can be easily linked with the malicious gadget as the brand's name is visible on the chips.

\includegraphics[scale=0.3]{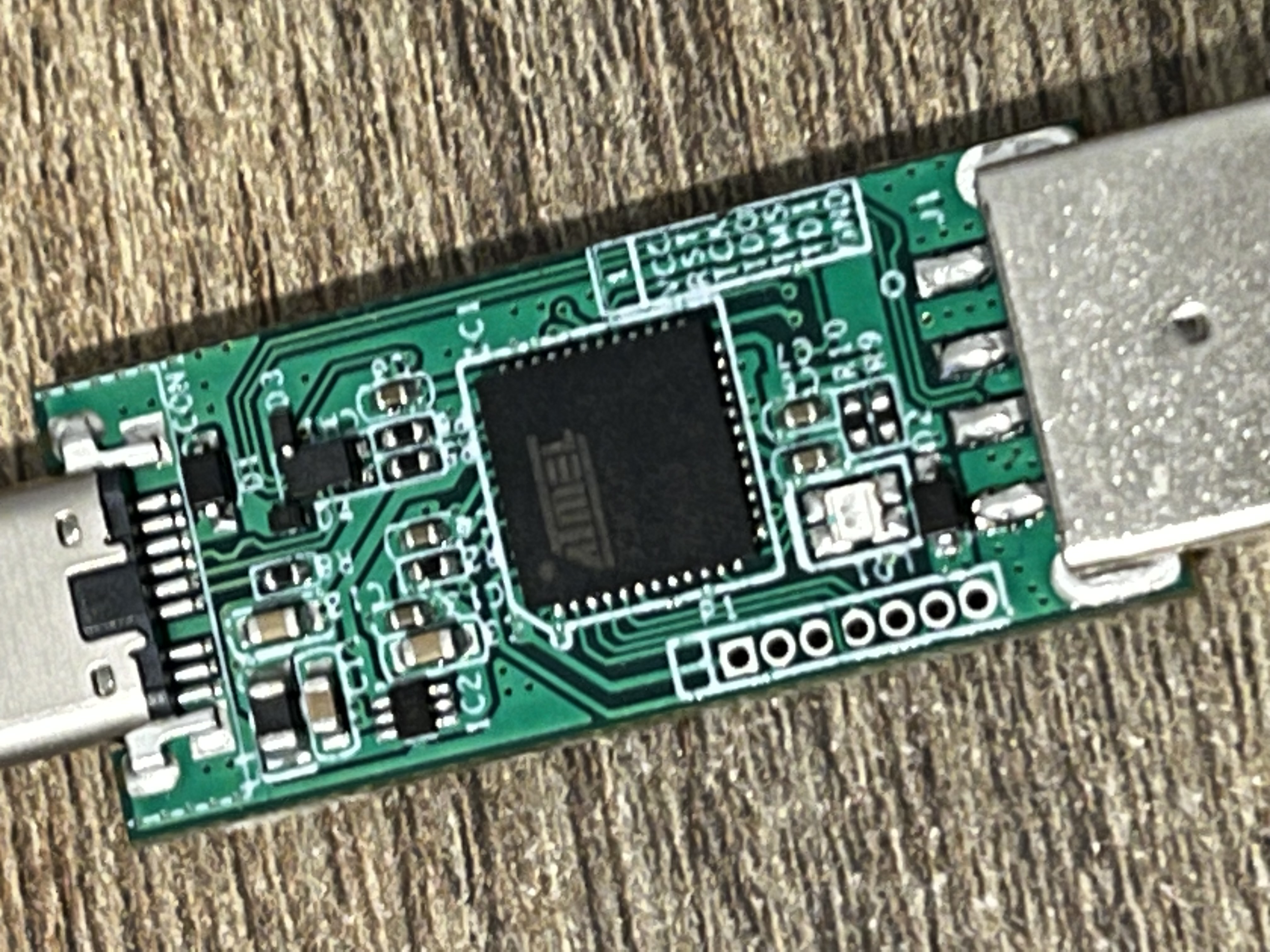}

Although, evading "hard-coded data"-based detections is out of scope for this study, it is worth noting that "Ruber Ducky" provides spoofing capabilities for Vendor and Product IDs. Given the behavioral nature of the detection and not basing our indicators on a sole kind of data, we can confidently say that even if such data is spoofed, we would still be able to identify forensic footprints.

Find below the relevant raw ETW-originating logs in chronological order with the timestamps from ETW converted to human-readable form:
\begin{lstlisting}
ImageLog:\Device\HarddiskVolume2\Windows\System32\drivers\hidusb.sys:EndImageLog

PnpLog:USBSTOR\Disk&Ven_ATMEL&Prod_Mass_Storage&Rev_1.00\7&85c08e4&0&111111111111&0:
2023-02-02 21:48:10.094 -08:00:EndPnpLog

ProcessLog:2023-02-02 21:48:17.752 -08:00:3740:
\Device\HarddiskVolume2\Windows\System32\WindowsPowerShell\v1.0\powershell.exe:EndProcessLog
ProcessLog:2023-02-02 21:48:17.849 -08:00:5248:
\Device\HarddiskVolume2\Windows\System32\conhost.exe:EndProcessLog
ProcessLog:2023-02-02 21:48:18.221 -08:00:8120:
\Device\HarddiskVolume2\Windows\System32\WindowsPowerShell\v1.0\powershell.exe:EndProcessLog
ProcessLog:2023-02-02 21:48:18.224 -08:00:4004:
\Device\HarddiskVolume2\Windows\System32\conhost.exe:EndProcessLog
\end{lstlisting}

Based on our timestamps, we can now search through the keystroke log and identify an average count to see if "powershell" launching was followed by fast-typing.

Based on a quick calculation, upon converting the time-stamps to Windows time, we identified \textbf{ a large amount of} keys to have been pressed within a span of \textbf{approximately 10 seconds} from just a moment prior to spawning "powershell", exceeding the average human typing capabilities. We could account the keystrokes pressed even before launching the command prompt of "PowerShell" but even through an approximation, the point has been proven and malicious BadUsb activity is highly probable to have happened based on this behavior only. If we regard the extra information from the device, we can secure our thoughts and claim to have successfully identified a BadUsb attack "post-mortem".

\section{Graphical Representation of Keystroke Peak}

In order to assist us get an idea of the actual spike and embrace a more friendly approach towards human cognitive capabilities, we will utilize a graph to represent the keystrokes across the line of time on a "mass scale".
Notice before the peak, the line of normal typing activity compared to when the "Ruber Ducky" was plugged in.
The blue graph can be safely compared to the ETW logs and represents the first forensic situation we are investigating. Please bare in mind that the keystroke number is an estimation and not a definitive number representing their amount.

\includegraphics[scale=0.5]{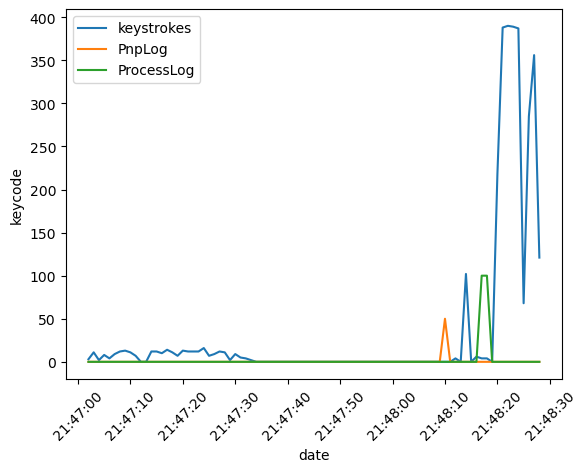}

\section{Conclusion}
This approach is by far \textbf{not} fool-proof and can increase in difficulty as the metrics increase, including the size of the logs and the attacker's attempts to obscure the timeline via sleeps and other strategies.
This approach could have been improved possibly with extra data visualization techniques, however the core would remain the same. The main idea exhibited is how someone can follow a "forensic trajectory-like" logic based on data from the system's kernel and tracing providers and how those data can be obtained safely and intuitively.

\section{References}
\url{https://www.bitdefender.com/blog/hotforsecurity/hp-laptops-found-carrying-keylogger-in-synaptics-touchpad-driver/}\newline
\url{https://shop.hak5.org/products/usb-rubber-ducky-textbook}\newline
\url{https://learn.microsoft.com/en-us/windows-hardware/drivers/ddi/kbdmou/ns-kbdmou-_connect_data}\newline
\url{https://github.com/microsoft/krabsetw/tree/master/examples/NativeExamples}\newline
\url{https://github.com/repnz/etw-providers-docs/blob/master/}\newline
\url{https://www.digital-detective.net/dcode/}\newline
\url{https://www.cyberpointllc.com/blog-posts/cp-logging-keystrokes-}\newline
\url{with-event-tracing-for-windows-etw.php}\newline
\end{document}